\documentclass[prstab,11pt,superscriptaddress,showpacs]{revtex4-1}

\usepackage{graphicx}
\usepackage{booktabs}
\usepackage{amssymb}
\usepackage{amsmath}
\usepackage{epstopdf}
\usepackage{hyperref}

\begin{document}

\title{Space charge compensation for intense ions beams in booster synchrotrons}

\author{O. Boine-Frankenheim}
\affiliation{Technische Universit\"at Darmstadt, Schlossgartenstr.8, 64289 Darmstadt, Germany}
\affiliation{GSI Helmholtzzentrum f\"ur Schwerionenforschung GmbH, Planckstr.1, 64291 Darmstadt, Germany}
\author{W. D. Stem}
\affiliation{Technische Universit\"at Darmstadt, Schlossgartenstr.8, 64289 Darmstadt, Germany}
\begin{abstract}
For booster synchrotrons, like the SIS at GSI, space charge is one of the main intensity limitations. 
At injection energy the space charge induced tune spreads in booster synchrotrons are large, 
reaching up to 0.5 and the ramping times are typically short. We study the effect of several localized electron lenses distributed equally around the circumference in order to compensate for the transverse space charge force.
\end{abstract}

\maketitle

\section{Introduction}
Space charge represents a major intensity limitation in booster synchrotrons operating at low or medium energies. The limitation arises due to stopbands caused by 
envelope instabilities driven by the lattice structure \cite{doi:10.1063/1.56781,Hofmann2017}
or due to the space charge induced tune spread and its overlap with incoherent, nonlinear resonances \cite{Franchetti2010}. Depending on the chosen working point and the intensity, coherent or incoherent effects can dominate or occur together. The space charge limit in booster synchrotrons is still an active field of research, both numerically as well as experimentally. 
After the success of electron lenses for beam-beam compensation in RHIC \cite{Gu2017}, the potential of such lenses to also compensate, at least partially, for space charge would increase the potential for higher intensities in booster synchrotrons. However, one should keep in mind that space charge acts as a distributed defocusing error, whereas the beam-beam tune shift has a local source. Therefore localized lenses cannot be expected to be as efficient for space charge as they are for beam-beam compensation. One has to install several lenses around the ring instead of placing one lens close to the beam-beam interaction section.  
Furthermore, in order to be effective for space charge, the electron current profile has to match the bunch profiles of the ions. For very short, relativistic bunches the optimum bunch overlap was studied in \cite{Litvinenko2014}. For booster synchrotrons with typically long bunches compared to the length of an electron lens, the matching of the electron current profile with the ion bunch profile might be easier to achieve in experiments.                  
Different concepts for space charge compensation were studied in \cite{Aiba2007}. 
Electron lenses for space charge compensation were also discussed in \cite{Shiltsev2016}.   The possibility of space charge compensation in booster synchrotrons was studied in \cite{Burov2000} using analytical models for the envelope instability and the coherent dipole tune. The limitation of the compensation degree by the upward shift of the coherent dipole tune was pointed out. For the FNAL booster synchrotron a minimum number of lenses     was estimated.     
For the SIS synchrotron, used as a reference machine in this study,
the incoherent resonances induced by the space charge field of the electron beam in the cooling section were analyzed in \cite{Sorge2007}. For electron cooling the tune shift induced by the electron beam is usually well below 0.1
and the ion beam currents are low. Therefore incoherent resonances dominate and were identified up to order 6
in the tracking studies together with a severe emittance growth.    
The present study relies on a simulation model for intense beams including 
the linear synchrotron lattice, the 2D self-consistent space charge force and simple kicks for the lenses. We compare the results of extended simulation scans to analytical models.   

\section{Space charge compensation: 2D model}

The compensation of space charge in bunches is a 3D problem. The transverse profile 
of the electron beams and its current profile both have to match
the transverse profile and the longitudinal bunch profile of the circulating ion bunches. 
In this study we assume ion bunches that are long compared to the interaction length with the electron beam. The current profile of the the electron beam in the interaction section is assumed 
to follow exactly the ion bunch profile.
If a certain space charge compensation degree is chosen for the bunch center, it will therefore hold also at the bunch ends. If we further assume that the relevant effects are faster than the synchrotron oscillation period, a 2D model for the bunch center slice is sufficient.    
The space charge tune shift in the bunch center is (vertical plane)
\begin{equation}
\Delta Q_y\approx-\frac{N Z^2 r_p}{2\pi\varepsilon_y \beta_0^2\gamma_0^3 A B_f},
\end{equation}
where $\beta_0$ and $\gamma_0$ are the relativistic parameters for the ion beam,
$N$ is the total particle number in the ring, $Z$ the charge state of the ions, $A$ the mass number, $B_f$ is the bunching factor, $\varepsilon_y$ is the unnormalized emittance of the equivalent KV beam. 

The beam-beam tune shift (co-propagating: $-$, counter-propagating: $+$) induced by one electron lens is
\begin{equation}
\Delta Q^e_y=\frac{1\mp\beta_e\beta_0}{\beta_e}\frac{ZI_elr_p}{2\pi Aec\varepsilon_y\beta_0^2\gamma_0}.
\end{equation}
Hereby we assume that the transverse profiles of both beams overlap ideally. $I_e$ is the current of the electron beam, $l$ is the length of the interaction section, $\beta_e$ is the velocity of the electrons divided by the speed of light. The required electron current in order to fully compensate a given space charge tune shift $\Delta Q_y$ with one lens is 
\begin{equation}
I_e=\frac{\beta_e}{1\mp\beta_e\beta_0}\frac{2\pi ecA\varepsilon\beta_0^2\gamma_0}{Zlr_p}\Delta Q_y.
\end{equation}
The degree of space charge compensation we define as
\begin{equation}
\alpha=\frac{\Delta Q^e}{\Delta Q},
\end{equation}
where $\Delta Q$ is the space charge tune shift and $\Delta Q^e$ the total beam-beam tune shift 
generated by $N_e$ electron lenses. 
The GSI SIS heavy-ion synchrotron \cite{Appel2016a} is used as a reference case for the study (see also \autoref{app:sis18}).
The SIS has a circumference of $C=216$ m and 
$S=12$ periodic sectors. The injection energy is 11.4 MeV/u. Example beam parameters are given 
in \autoref{tab:sis}. The parameters of the existing SIS electron cooler are given in \autoref{tab:cool}.
\begin{table}[h]
\begin{tabular}{| l | l |}
\hline
Ion & Ar$^{18+}$ \\ \hline
$E_0$ & 11.4 MeV/u \\ \hline
$N_b$ & $5\times 10^{10}$ \\ \hline
$h$ & 2 \\ \hline
$B_f$ & 0.3 \\ \hline
$\varepsilon_{x,y}$ & 150/50 mm mrad \\ \hline
$\delta p/p$ & $10^{-3}$ \\ \hline
-$\Delta Q_{x,y}$ & 0.2/0.5 \\ \hline
$Q_{x,y}$ & 4.14/3.6 \\ \hline
\end{tabular}
\caption{Reference beam parameters for the SIS at injection energy.
$B_f$ is the bunching factor. $N_b$ the number of ions per bunch. $h$ is the harmonic number (number of bunches).$\varepsilon_{x,y}$ are the absolute emittances ($4\times$ the rms emittances). $Q_{x,y}$ are the bare reference working points.}
\label{tab:sis}
\end{table}
\begin{table}[h]
\begin{tabular}{| l | l |}
\hline
Max. electron energy & 35 keV \\ \hline
Max. current & 2 A \\ \hline
effective length & 2.8 m \\ \hline
$\beta_{x,y}$   & 8/15 m \\ \hline
\end{tabular}
\caption{SIS electron cooler parameter. The $\beta$ functions are the average values in the cooler section.}
\label{tab:cool}
\end{table}
As a simple example, if we use the SIS cooler in the conventional way (co-propagating, $\beta_e=\beta_0$)
and assume a symmetric Ar$^{18+}$ beam with $\varepsilon_y=20$ mmrad a tune shift 
of $\Delta Q_y=-0.1$ could be compensated (see also \cite{Sorge2007}). This of course requires an ideal overlap of both beams. The compensation of the reference beam parameters given in \autoref{tab:sis} 
would require at least an electron current of 2 A and $N_e=3$. 

\section{Stability considerations: Orbit and Optical functions }\label{sec:stability}


The transverse space charge force acts as a distributed (de)focusing error around the ring. In periodic focusing lattices space charge alone only leads to incoherent structural or coherent parametric resonances. 
On the contrary the electron lenses act as localized focusing errors. If they are spaced symmetrically around the synchrotron they define new periodic structure cells (for $N_e<S$). The lenses also result in additional nonlinear error resonances. 

Before discussing discreteness effects, the limitation of the compensation degree due to the shift of the coherent tune is discussed. This limitation has been pointed out in \cite{Burov2000}.  
The focusing electron lenses cause an upward shift of the coherent tune 
(the frequency of the beam centroid oscillations)
\begin{equation}
Q_c=Q_0+\alpha\Delta Q,
\end{equation}
where $Q_c$ is the coherent tune and $Q_0$ the bare tune. This has been experimentally shown at the SIS for both coasting and bunched beams in \cite{Stem2016}. An important limitation for the total strength of the lenses
is that $Q_c$ should stay well below integers $n$, in order
for the closed orbit to remain stable. This limitation 
only depends on the total beam-beam tune shift and not on the number of lenses.

The number of lenses $N_e$ is important for the perturbation 
of the optical functions. The beating of the $\beta$-function caused by an electron lens $j$, treated here 
as a localized focusing error is (see e.g.\cite{Lee1999})
\begin{equation}
\frac{\Delta\beta(s)}{\beta(s)}=\frac{2\pi\Delta Q^e_j}{\sin(2\pi Q/N_e)}\cos(2|\phi_j-\phi|-2\pi Q/N_e).
\end{equation}
Assuming that the lenses are located symmetrically around the ring,  
$\phi_j$ is the phase advance from the beginning of the section 
to the position of the lens and $\phi$ the phase advance at the observation point.
The maximum beta beating caused by the lenses is approximately
\begin{equation}
\left(\frac{\Delta\beta}{\beta}\right)_{\max}\approx 2\pi\Delta Q^e / N_e,
\end{equation}
where $\Delta Q^e$ is the total beam-beam 
tune shift.
If we require that the maximum beta beating remains well below 1,
only very small space charge tune shifts can be compensated
with one lens (well below 0.1 in the SIS). For the typical space charge tune shifts 
in high current synchrotrons, of the order of $0.4-0.5$, 
several (at least $N_e\gtrsim 3$ for the SIS) lenses are required to at least partially 
compensate the space charge tune shift, without causing a large beating 
amplitude of the optical functions (see also \autoref{app:sis18} for more detailed results). 

It is important to notice that for dc electron beams the beta beating could be reduced by corrector quadrupoles. However, for the desired pulsed electron beam operation, a correction with fixed gradient quadrupoles is not possible.

\begin{figure}[htb]
    \centering
    \includegraphics[width=90mm]{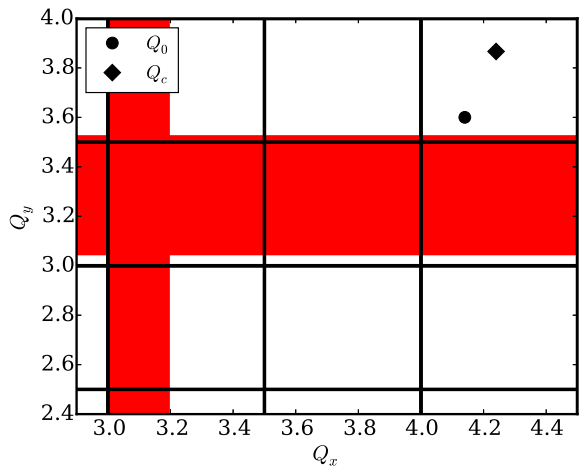}
    \caption{Stopband (red) from \autoref{eq:stability} for $N_e=3$, $\Delta Q=0.8$ and a compensation degree $\alpha=0.33$. Indicated is also the reference high-current working point in SIS $Q_0$ together with the resulting upwards shifted coherent tune $Q_c$.}
    \label{fig:stability}
 \end{figure}

The required number of lenses can also be estimated from the 
stability criteria for betatron oscillations in periodic focusing lattices:
\begin{equation}\label{eq:stability}
|\cos(2\pi Q/N_e)-2\pi\frac{\Delta Q^e}{N_e}\sin(2\pi Q/N_e)|<1,
\end{equation}
where the tune $Q$ is the particle tune and $\Delta Q^e$ is the total beam-beam tune shift induced by the $N_e$ electron lenses. 
The area in tune space with $|\ldots|\geq 1$ are stopbands in which the amplitude of linear betatron oscillations grows exponentially. Including the space charge tune shift in the particle tune $Q=Q_0+\Delta Q$ causes a 
shift of the stopbands, but leaves the stopband width unchanged. 
The stopband width is determined by the beam-beam tune shift and the density of bands by $N_e$. An example case
for the SIS with $N_e=3$ lenses, an assumed enlarged space charge tune shift of $\Delta Q_y=-0.8$ ($\Delta Q_x=-0.2$) and a compensation degree $\alpha=0.33$ is shown in \autoref{fig:stability}. The distance between stopbands is $N_e/2$, which is also why we call those stopbands "$180^\textrm{o}$". 
The stopband width increases as $\delta Q \approx 2 \Delta Q^e$. \autoref{fig:stability} indicates that for $N_e=3$ a total beam-beam tune shift of $\Delta Q^e_y=0.2-0.3$ might be feasible. 
The degree of compensation is limited by the coherent tune
approaching the integer value $Q_c=4$. How close $Q_c$ is allowed to approach 
the integer resonance depends also on the accuracy of the alignment of the electron lenses and the circulating ion beam. The tolerance for such offsets and the resulting localized dipole kicks are not part of this study.      

\section{Envelope perturbations and instabilities with electron lenses}\label{sec:coh}

The 2D envelope equations (see for example \cite{Yuan2017}) can be solved numerically 
including the electron lenses, treated as thin, linear defocusing elements. 
Thereby one can obtain space charge structural instabilities 
and their modification due to the lenses. Furthermore, for stable beams, 
this self-consistent model can predict the perturbation of the optical functions 
and, for unstable beams, the stopbands with localized lenses and linear space charge.  

First, we use simple analytical expressions to discuss the expected modifications and shifts of 
space charge structural resonances due to the electron lenses.  
The condition for a parametric resonance or envelope instability including a partial incoherent space charge compensation is      
\begin{equation}\label{eq:parares}
2(Q_0 - \alpha\Delta Q) - \Delta Q_2 = \frac{n}{2}S
\end{equation}
where $Q_0$ is the bare tune, $\Delta Q_2$ is the coherent tune shift 
for the envelope modes, $n$ is the harmonic number and $S=N_e$ is the number of structure 
cells, which equals the number of electron lenses in our scenario. 
It is important to note that the stopbands arising from (coherent) envelope instabilities, 
also called $90^\textrm{o}$ stopbands, are different from the $180^\textrm{o}$ stopbands shown in \autoref{sec:stability}.       

Both stopbands can overlap, as it will be the case in the SIS close to $Q_y=3$. 
In this case the $180^\textrm{o}$ stopband dominates, as it defines the linear stability for the 
single particle and envelope betatron oscillations.

The coherent tune shifts in \autoref{eq:parares} can be estimated
in $x$ and $y$ assuming uncoupled envelope oscillations \cite{Boine-Frankenheim2016}
\begin{equation}\begin{split}
\Delta Q_2^x=\left(1-\frac{1+2\eta_0}{4(1+\eta_0)}\right)\Delta Q_x ~\\
\Delta Q_2^y=\left(1-\frac{2+\eta_0}{4(1+\eta_0)}\right)\Delta Q_y
\end{split}
\label{eq:collective tunes}
\end{equation}
With $a$ and $b$ the beam radii in $x$ and $y$ for vanishing space charge, the ellipticity parameter $\eta_0$ is given for equal emittances as $\eta_0 \equiv{a}/{b} =\sqrt{Q_{0,y}/Q_{0,x}}$. 

The coherent tune shift $\Delta Q_2$ is smaller (factor 5/8 for a symmetric beam) 
than the incoherent tune shift $\Delta Q$. This effect is often 
called "coherent advantage": With increasing space charge the coherent envelope tune will hit 
the resonance after the incoherent particle tune. For strong space 
charge, like in booster rings, the relevant resonance is assumed to be the coherent and not the 
incoherent one \cite{Hofmann2017}.

The stopbands corresponding to the envelope instability are shifted 
downwards by the presence of the electron lenses. Because of 
the "coherent advantage" the compensation can be partial 
$\Delta Q^e=\alpha \Delta Q$, with $\alpha\approx 5/8$ for symmetric beams.  
 
\begin{figure}[htb]
    \centering
    \includegraphics[width=90mm]{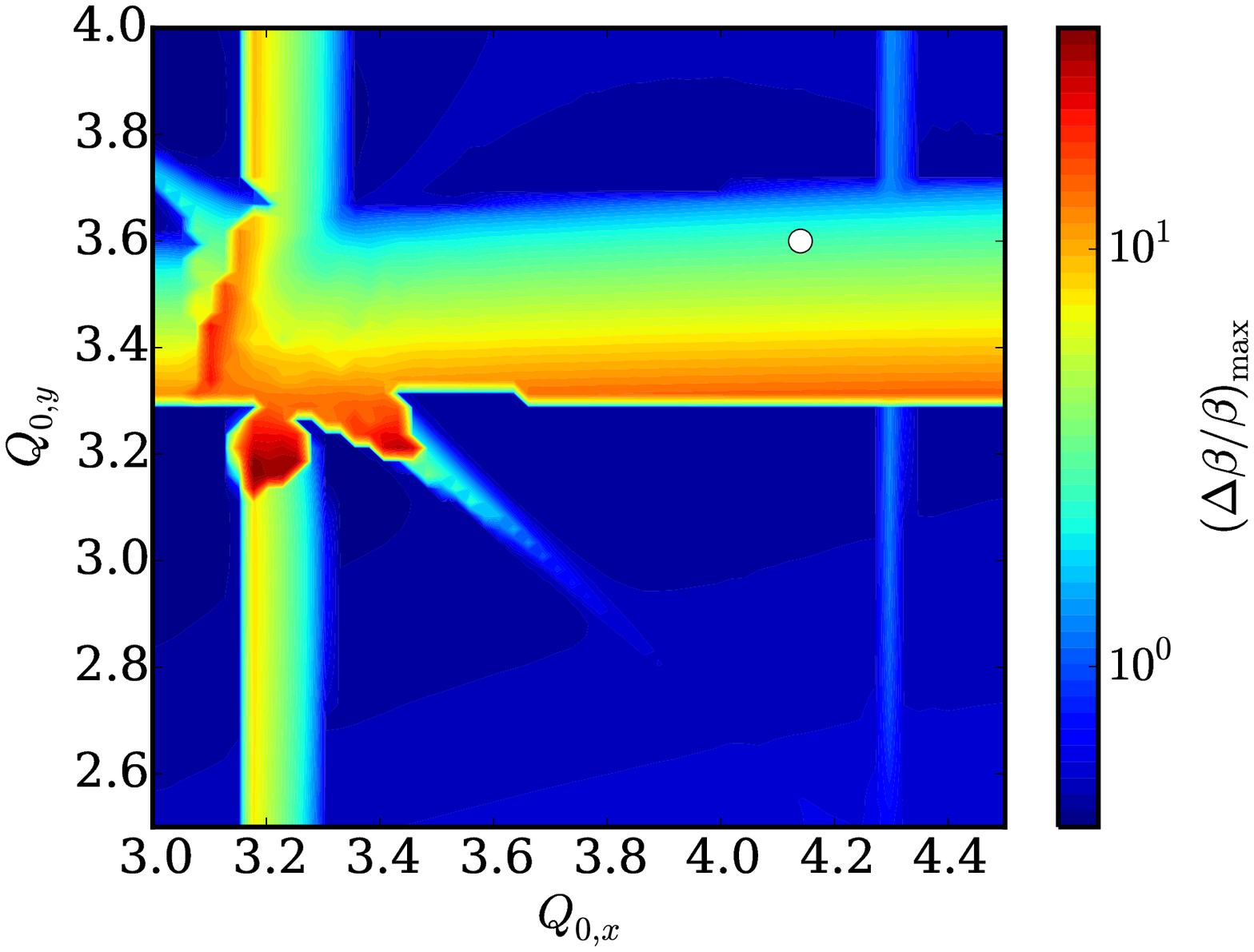}
     \includegraphics[width=90mm]{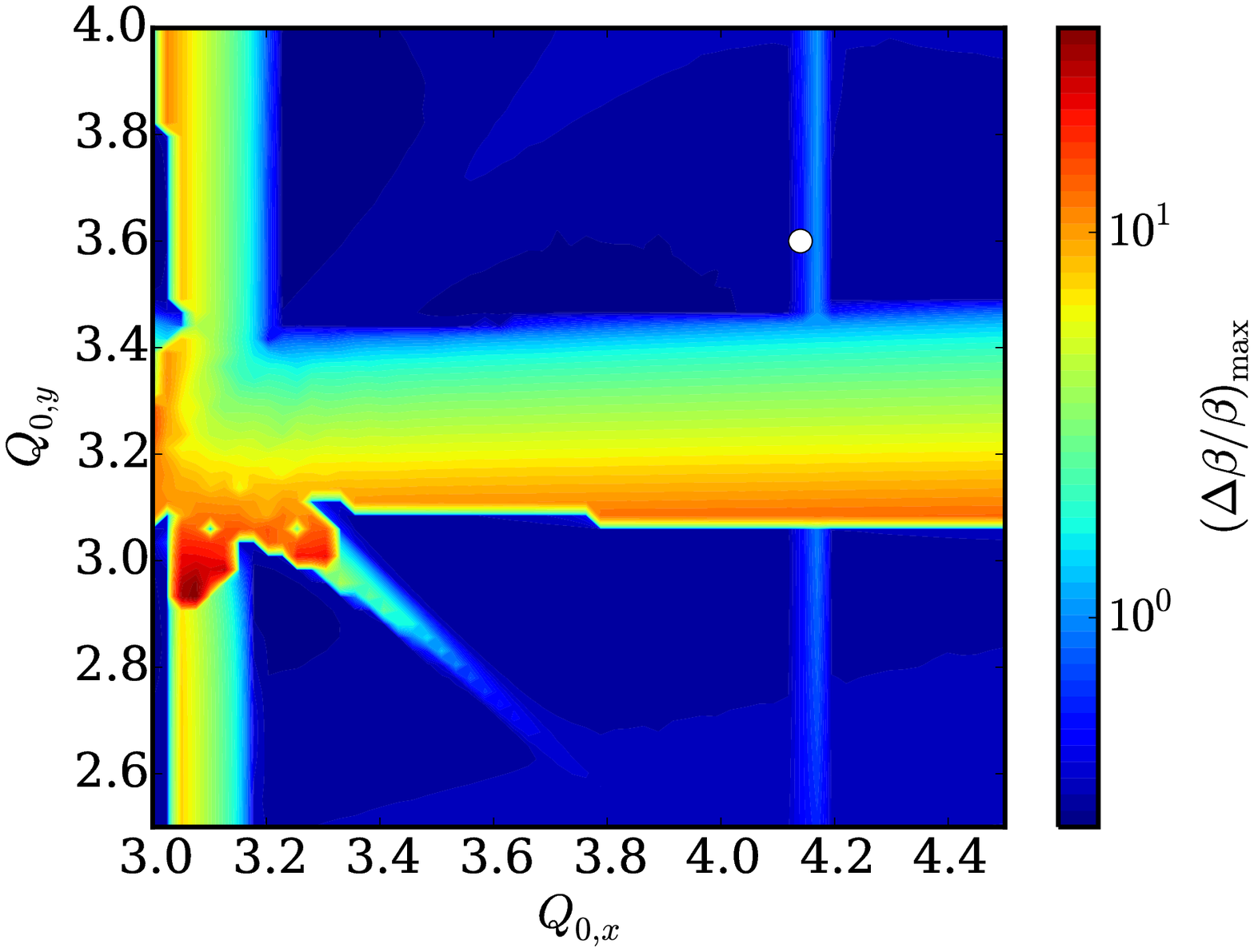}
    \caption{Envelope simulation scans for $N_e=0$ (upper) and $N_e=S$ (lower plot,$\alpha=0.33$) with initial tune shifts
     $\Delta Q_{x,y}=-0.3/0.8$. The reference working point is indicated as a white dot.}
    \label{fig:env_N0_dq08}
 \end{figure}

Next we solve the time-dependent 2D envelope equations numerically
and plot the results for the maximum beta function mismatch amplitude $\Delta\beta/\beta$ as a function of the bare tunes. 
The result of such tune scans are shown in \autoref{fig:env_N0_dq08}.
For the scan initial tune shifts $\Delta Q_{x,y}=-0.3/0.8$, which are higher than the 
SIS reference parameters. The scan shown on upper plot is performed without electron lenses. The 
shift of the lower edge of the two stopbands (in $x$ and $y$) is well described 
by \autoref{eq:parares} with $\alpha=0$ and $S=12$. The white dot indicates the reference high-current 
working point in the SIS, which lies inside the unstable area for the chosen beam intensity.  
Very weakly visible are the parametric sum resonance (the diagonal line) \cite{Boine-Frankenheim2016} and the dispersion 
induced envelope resonance (the vertical line at $Q_x=4.3$) \cite{Yuan2017}.

The lower part of \autoref{fig:env_N0_dq08} shows a tune scan performed 
with $N_e=12$ electron lenses (one in each SIS lattice period). The lenses 
are adjusted to provide a partial compensation of the space charge tune shift by one-third ($\alpha=0.33$), limited by the integer resonance. The stopband's lower edge, which is again well described by \autoref{eq:parares}, remains close to $Q_y=3$. The width of the stopbands remains unaffected by the presence of the lenses. 
The effect of the $N_e=S$ lenses is basically only to shift the $90^\textrm{o}$ stopbands downwards.

\begin{figure}[htb]
    \centering
    \includegraphics[width=90mm]{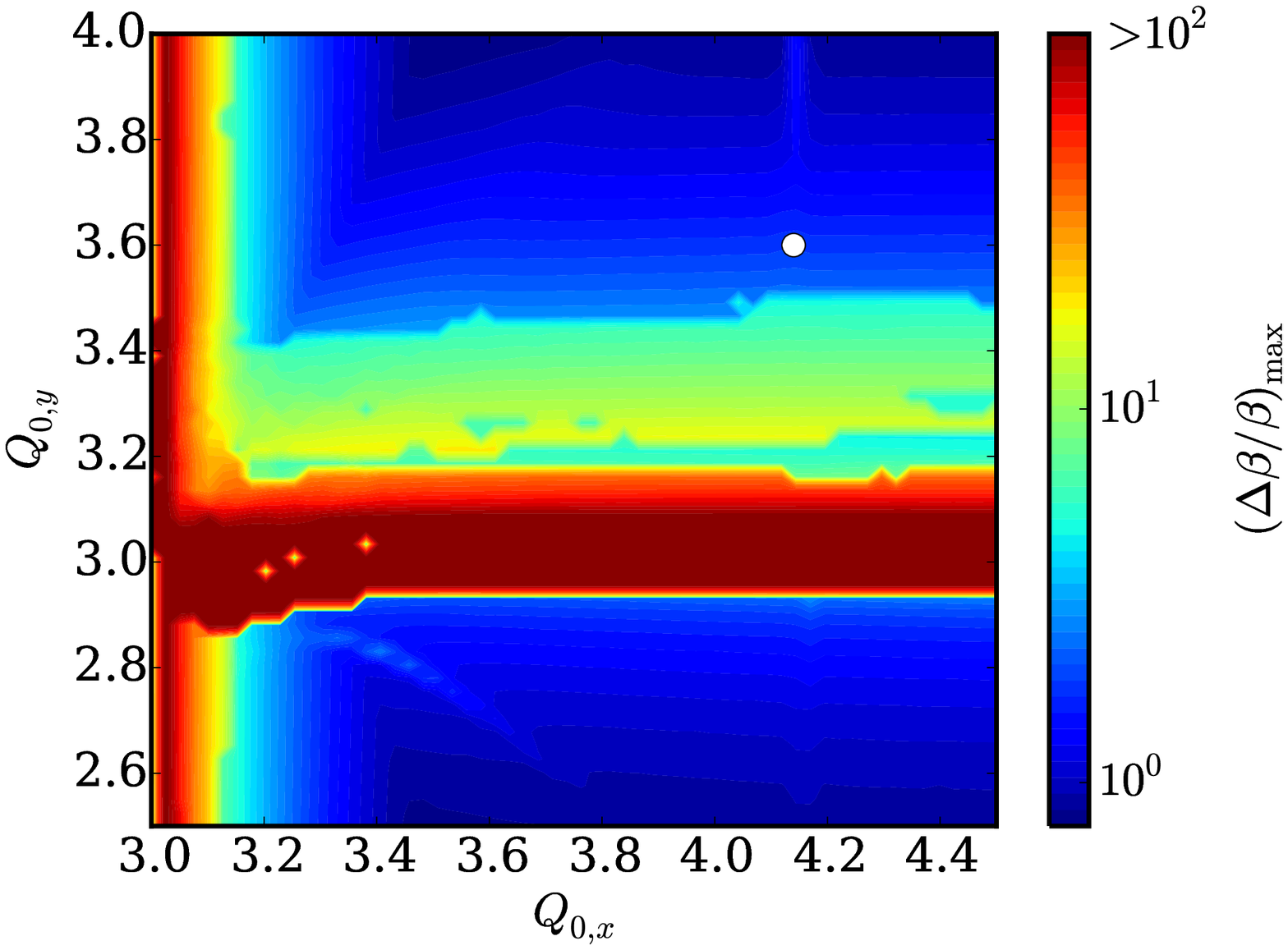}
    \caption{Envelope simulation scan for $N_e=6$ and $\Delta Q_{x,y}=-0.3/0.8$ and $\alpha=0.33$.}
    \label{fig:env_N6_dq08}
 \end{figure}

If we reduce the number of lenses $N_e$ to six (and so also the periodicity of the ring), keeping the compensation at $\alpha=0.33$, the resulting scan is shown in \autoref{fig:env_N6_dq08}. The stopband close to $Q_v=3$ now 
broadens and resembles the $180^\textrm{o}$ stopband shown in \autoref{fig:stability}. Furthermore the stable beam (blue areas) shows strong beta beating. For $N_e=3$ (not shown) the situation worsens and stable areas with tolerable beta beating are difficult to identify.  

\section{Particle-In-Cell (PIC) tracking: incoherent and coherent resonances with electron lenses}

\begin{figure}[htb]
    \centering
    \includegraphics[width=90mm]{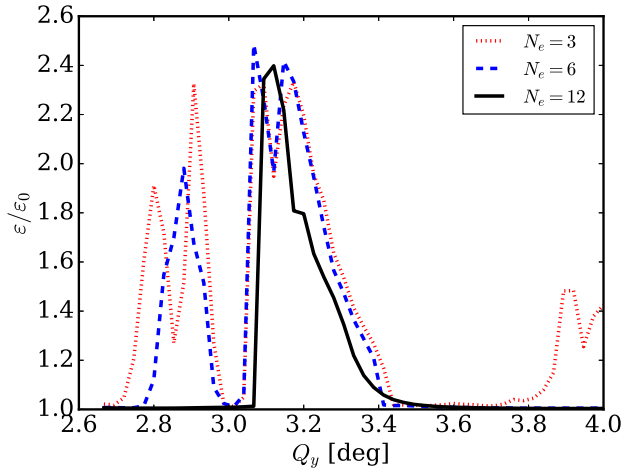}
    \caption{PIC simulation scan for $N_e=3,6,12$, $\Delta Q_{x,y}=-0.3/0.8$ and $\alpha=0.33$.}
    \label{fig:pic_alpha03_dq08}
 \end{figure}

The 2D envelope model, discussed in the previous section, allows very fast tune 
scans and helps to separate coherent and incoherent effects, which occur together 
in Particle-In-Cell (PIC) tracking simulations. In our simplified model incoherent resonances are excited by the (static) nonlinear beam-beam forces and by the $n=0$ harmonic of the nonlinear space charge force.  We use an initial waterbag distribution for the PIC macro-particles, which might be closer to the actual 
distribution of the injected beam in a booster synchrotron than a Gaussian. 
A tune scan along a vertical line through the reference SIS working point performed with the PIC simulation code 
pyPATRIC is shown in \autoref{fig:pic_alpha03_dq08}. The beam and compensation parameters are the same as the ones used in the previous section. The emittance growth observed after $500$ cells is shown for $N_e=3,6,12$ lenses. For $N_e=12$ the envelope stopband with its peak close to $Q_y=3.1$ is present also in the PIC simulations. For $N_e=6$ and $N_e=3$ between 
$Q_y=3.0$ and $Q_y=3.4$, in agreement with the stability condition \autoref{eq:stability}, the broader $180^\textrm{o}$ stopband appears
together with presumably fourth order resonances below $Q_y=3$ and $Q_y=4$ induced by the nonlinear lenses. 

\begin{figure}[htb]
    \centering
    \includegraphics[width=90mm]{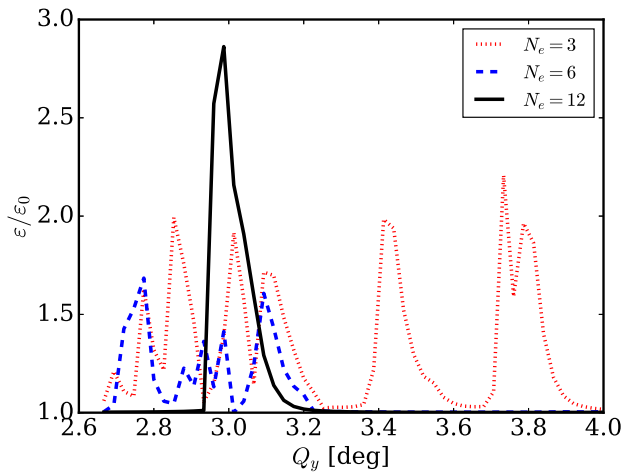}
    \caption{PIC simulation scan for $N_e=3,6,12$, $\Delta Q_{x,y}=-0.2/0.5$ and $\alpha=0.5$.}
    \label{fig:pic_alpha05_dq05}
 \end{figure}

For the less intense SIS reference beam \autoref{tab:sis} the compensation, limited by the integer resonance, can be larger $\alpha\approx 0.5$. 
The beam-beam tune shift $\Delta Q^e\approx 0.25$ in this case is still similar to the more intense beam case. The result of the scan for the reference beam is shown in \autoref{fig:pic_alpha05_dq05}. For $N_e=12$ the envelope stopband is now close to $Q_y=3$. For $N_e=6$ and $N_e=3$ the $180^\textrm{o}$ stopband appears between 
$Q_y=2.7$ and $Q_y=3.2$. For $N_e=3$ two presumably fourth order resonance lines appear above the $180^\textrm{o}$ stopband.  

\begin{figure}[htb]
    \centering
    \includegraphics[width=90mm]{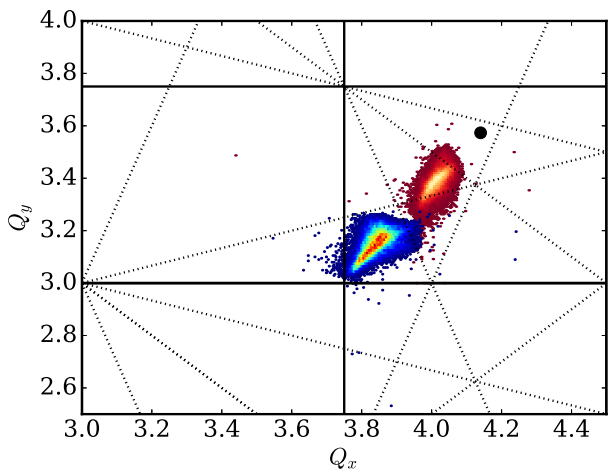}
    \caption{Tune footprint (red) for $N_e=3$, $\Delta Q_{x,y}=-0.2/0.5$ and $\alpha=0.5$. For comparison, the tune footprint for the uncompensated beam is shown as well (red/blue). The fourth order resonances excited by $N_e=3$ lenses are shown.}
    \label{fig:tune_WB_N3_dq05}
 \end{figure}

An example resonance diagram (dashed fourth order lines) and tune footprint for $N_e=3$ 
is shown in \autoref{fig:tune_WB_N3_dq05} for $\Delta Q_{x,y}=-0.2/0.5$ and $\alpha=0.5$ (red). For comparison, the tune footprint for the uncompensated beam is shown as well (red/blue).

It is important to note that in our 2D PIC model we use electron lenses that are 
ideally matched to the ion beam. In more realistic scenarios the electron beam would
be kept strictly round, because the beam is guided by a solenoid field 
and affected by the $E\times B$ rotation (see also \autoref{app:sis18}).
We found that for round electrons beams different nonlinear resonances are excited, compared to the fully elliptical beams used in this model.  

\section{Conclusions and Outlook}

Within a simplified 2D beam dynamics model we studied the (partial) compensation of large space charge tune shifts, typical for booster synchrotrons, by localized electron lenses. The model includes self-consistent 2D space charge and fixed (nonlinear) kicks from the localized electron lenses. We use the GSI SIS18 heavy-ion synchrotron as reference case and only account for transverse space charge and the lenses as the only 'error' sources in our simulation model.  
From the envelope model we obtain the beta beating amplitudes, the $180^\textrm{o}$ stopbands 
(caused by the localized gradient errors) as well as the envelope instability or $90^\textrm{o}$ stopbands (caused by space charge structural instabilities). The PIC simulation provides additional information on nonlinear resonances excited by the lenses, which cannot be resolved in the envelope model.

The degree of space charge compensation is limited by the integer resonance and orbit stability. This limitation 
has been assumed here in order to adjust the compensation degree.

If the number of lenses $N_e$ equals the lattice periodicity ($S=12$), we find that the lenses shift the $90^\textrm{o}$ stopbands downwards, without affecting their width. In this case the $180^\textrm{o}$ stopband are not relevant, as the phase advance per cell is usually chosen well below $180^\textrm{o}$.   
For $N_e < S$ the localized lenses reduce the symmetry of the lattice     
and cause severe beta beating as well as $180^\textrm{o}$ stopbands, spaced every $N_e/2$ in 
tune space. Their width corresponds to the total beam-beam tune shift induced by the lenses and their center position is shifted by space charge. The $180^\textrm{o}$ stopband usually overlaps and substitutes a $90^\textrm{o}$ stopband. Furthermore, incoherent nonlinear error resonances induced by the lenses are clearly visible in the PIC simulation scans.      

In conclusion, for the large space charge tune shifts typical for booster synchrotrons a partial compensation by $\Delta Q=0.2-0.3$ might be possible, provided that there are sufficiently many lenses. For the SIS we estimate a minimum amount of $N_e=6$ in order to have a reasonable chance to find a working point range that allows for partial compensation, outside of any stopbands and nonlinear resonances induced by the lenses and with tolerable beta beating.        
For any $N_e<S$ the unwanted effects caused by the lenses seem to dominate in our example study. One has to keep in mind that in our model many additional sources of errors are not included, for example systematic and random errors in the alignment of the lenses and in the 3D overlap between the ion and electron beams. The alignment errors of the lenses will add to the closed orbit instability at integer coherent tunes. The errors in the electron current for different lenses will lead to additional 
gradient errors stopbands. Also we did not include any lattice errors, which are usually the reason for the space charge limit without the lenses. Therefore we expect our model to be very optimistic.           

Still there is room for further studies and possible improvements. For example, instead of trying to ideally overlap both beams one could use a transverse McMillan lens profile \cite{Stancari2015}. This could reduce the effect of nonlinear resonances induced by the lenses, but would not affect the error stopbands or the beta beating. The compensation might work better for lower space charge tune shifts ($0.1$ or lower) and a space charge limit dominated by nonlinear lattice errors. 
This would require studies with a 3D simulation model including synchrotron motion over much longer time scales than in this study. We plan to conduct such long-term simulation studies for the FAIR SIS100 synchrotron, first within a frozen 3D space charge model.    

\appendix

\section{Perturbation of the optical functions in the SIS}\label{app:sis18}
\begin{figure}[htb]
    \centering
    \includegraphics[width=90mm]{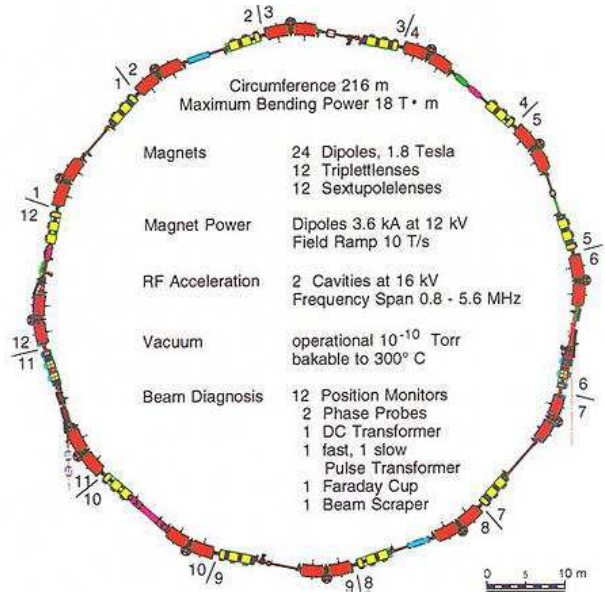}
    \caption{The SIS18 synchrotron. The existing electron cooler is localized in section 10.}
    \label{fig:sisring}
 \end{figure}

We assume that all electron lenses in the SIS will be located at the same position as the existing electron cooler in section 10 (see \autoref{fig:sisring}).  

\begin{figure}[htb]
    \centering
    \includegraphics[width=90mm]{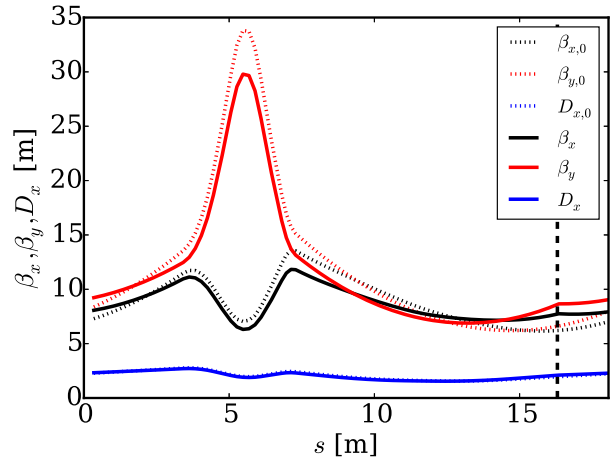}
    \caption{The optical functions in the one SIS cell. Shown are the $\beta$ function and the dispersion $D$ without space charge and lens (index $0$, dotted curves) and the ones for space charge tunes shifts $-\Delta Q_{x,y}=0.5/0.2$ and full compensation ($\alpha=1$) by one thin lens in every cell (lens location indicated by dashed vertical line).}
    \label{fig:sisN12}
 \end{figure}

\begin{figure}[htb]
    \centering
    \includegraphics[width=90mm]{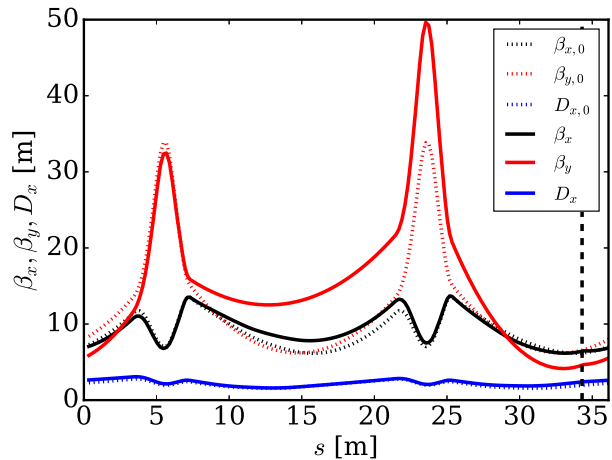}
    \caption{The optical functions in the two SIS18 cells. Shown are the $\beta$ function and the dispersion $D$ without space charge and lens (index $0$, dotted curves) and the ones for space charge tunes shifts $-\Delta Q_{x,y}=0.5/0.2$
    and half compensation ($\alpha=0.5$) by one thin lens in every second cell (lens location indicated by dashed vertical line).}
    \label{fig:sisN6}
 \end{figure}

The optical functions including space charge and the thin electron lens for one SIS cell obtained within our envelope model (not all lattice elements are included) are shown in \autoref{fig:sisN12} for $N_e=12$ and in \autoref{fig:sisN6} for $N_e=6$.


It is important to note that the ion beam at the location of the lenses will in general not be round. 
Even with equal $\beta$ functions in $x,y$ the asymmetric emittances will lead to an elliptical beam profile at the location of the lens. The $\beta$ beating can further enhance the difference in the $\beta$ functions. The electron beam guided by a solenoid field and affected by the $E\times B$ rotation, on the other hand, is usually round. This further complicates the matching of the transverse profiles of both beams. In our simplified model we treat both beams as ideally matched.
Furthermore the phase advance along the interaction sections is finite, whereas we use a thin 
lens approximation. 


\bibliographystyle{aipnum4-1}   
\bibliography{sc-comp-concept}

\end{document}